\documentclass{PoS}
\usepackage[]{latexsym}
\usepackage[]{amsfonts}
\usepackage[]{amsmath}

\newcommand{\Tr}{\textrm{Tr}}

\title{Eguchi-Kawai model with dynamical adjoint fermions}

\ShortTitle{Eguchi-Kawai model with dynamical adjoint fermions}

\author{\speaker{Ari Hietanen}\\
        Florida International University\\
        E-mail: \email{ari.hietanen@fiu.edu}}

\author{Rajamana Narayanan\\
        Florida International University\\
        E-mail: \email{rajamani.narayanan@fiu.edu}}

\abstract{It is believed that fermions in adjoint representation on
  single site lattice will  restore the center symmetry, which is a
  crucial requirement for the volume independence of large-N 
lattice gauge theories. We present a perturbative analysis which
supports the assumption for overlap fermions, but shows that center
symmetry is broken for naive fermions.}

\FullConference{The XXVII International Symposium on Lattice Field Theory - LAT2009\\
		 July 26-31 2009\\
		 Peking University, Beijing, China}

\begin{document}

\section{Introduction}
Eguchi and Kawai proposed \cite{Eguchi:1982nm} that a pure gauge
SU($N$) lattice theory can be reduced to a single site at the limit
$N\rightarrow\infty$. The proof was based on the assumption that the
single site has  a center symmetry associated with the traces of
the Polyakov loops. However, later the center symmetry was shown to be
broken in the continuum limit \cite{Bhanot:1982sh}.

It was immediately clear that EK-reduction might work for theories
with supersymmetry \cite{Mkrtchyan}. This has been explored recently
in the continuum \cite{Kovtun:2007py}, where it was 
suggested that fermions obeying periodic boundary
conditions in the adjoint representation, QCD(Adj), would be volume
independent. A continuum perturbation theory analysis on  ${\bf
  R}^3\times S^1$ showed $Z_N$ symmetry to be restored. Similar
calculations on $S^3\times S^1$ also showed a center symmetric phase
\cite{Hollowood:2009sy}. A perturbation theory analysis on lattice
with Wilson fermions and with one
small direction agreed with the result
\cite{Bringoltz:2009mi,Bringoltz:2009fj} (see also
\cite{Bedaque:2009md}). The perturbation theory results have also been
confirmed by numerical studies on a single site lattice
\cite{Bringoltz:2009kb}.

In this work we describe a perturbative calculation of  QCD(Adj) on a
single site lattice with na\"ive and overlap fermions. The results
show that the center symmetry is not restored for na\"ive fermions but
it is restored for overlap fermions for a wide range of Wilson mass
values $m$. The study of na\"ive fermions is motivated since we do not
expect doublers on a single site lattice with adjoint fermions. The
complete study including lattice simulations is presented in \cite{Hietanen:2009ex}. 

\section{Eguchi-Kawai model with adjoint fermions}
Eguchi-Kawai model with fermions is defined with the action
\begin{equation}
  S = S^g + S^f,
\end{equation}
where
\begin{equation}
  S^g = -bN \sum_{\mu\ne\nu=1}^4 \Tr  \left [ U_\mu U_\nu U_\mu^\dagger U_\nu^\dagger\right]
\end{equation}
and
\begin{equation}
  S^f =  -f \log \det H_{n,o} = -f \Tr \log H_{n,o}\label{sno},
\end{equation}
where $U_\mu\in$ SU($N$) are the link matrices,  $b=\frac{1}{g^2N}$ is
the gauge coupling constant, $H_{n,o}$ is the naive or overlap fermion
operator, and $f$ is the number of Dirac fermion flavors. 

Matrices $V_\mu$ are the link matrices in adjoint representation
calculated from $U_\mu$ as 
\begin{equation}
   V^{ab}=\frac12\Tr[T^aUT^bU^\dagger],
\end{equation}
where $T^a$ are the hermitian generators of SU(N) normalized as
$\Tr[T^a T^b] = 2\delta^{ab}$. 

Both $H_n$ and $H_o$ are $4(N^2-1)\times 4(N^2-1)$ hermitian matrices
and correspond to na\"ive Dirac fermions and overlap Dirac fermions
respectively. The exact forms are given in \cite{Hietanen:2009ex}.
The determinant of $H_{n,o}$ is positive definite and therefore the
logarithm is well defined. Furthermore, we can define a
hermitian matrix $\Sigma$ 
\begin{align}
  \Sigma = \begin{pmatrix}\sigma_2 & 0 \cr 0 & -\sigma_2\cr \end{pmatrix};\\ 
  \Sigma^\dagger = \Sigma;\ \ \ \Sigma^2=1,
  \label{sigdef}
\end{align}
such that,
\begin{equation}
\Sigma H_{n,o} \Sigma = H^*_{n,o}\label{hadjiden}
\end{equation}
which implies that all eigenvalues of $H_{n,o}$ are doubly degenerate
reflecting the adjoint nature of the fermions.
In addition, both na\"ive and overlap fermions obey chiral symmetry
and therefore the eigenvalues of $H_{n,o}$ will come in $\pm$ pairs.
Therefore the factor, $f$, in front of $S^f$ can be an integer (single Dirac
flavor)  or half-integer (single Majorana flavor) for all values of
$N$.\footnote{Note that one should have written $\gamma_5H_{n,o}$ in
(\ref{sno}) but this is the same as writing $H_{n,o}$ as long as $f$ is
an integer multiple of $\frac{1}{2}$.}

\section{Weak coupling expansion}\label{wca}
Our aim is to find out if the $Z_N^4$ symmetries are broken in the
weak coupling limit. We 
follow~\cite{Bhanot:1982sh} and perform the weak coupling
analysis by decomposing $U_\mu$ according to
\begin{equation} U_\mu = e^{ia_\mu} D_\mu e^{-ia_\mu};\ \ \ \ 
D_\mu^{ij}=e^{i\theta_\mu^i}\delta_{ij}. 
\end{equation}
Keeping $\theta_\mu^i$ fixed, we expand in powers of
$a_\mu$.
The lowest contribution to $S_g$ comes from the quadratic term in
$a_\mu$~\cite{Bhanot:1982sh} and the lowest contribution
to $S_f$ comes from setting $a_\mu=0$.

The computation of the fermion determinant
reduces to a free field calculation at this order
and the result is
\begin{equation} 
S_{n,o} = -4f \sum_{i\ne j} \ln \lambda_{n,o}( \theta^i-\theta^j+\phi)-4(N-1)f\ln\lambda_{n,o}(\phi),
\end{equation} 
where $e^{i\phi_\mu}$ is the boundary condition in the $\mu$ direction.
The eigenvalues, $\pm\lambda(p)$, are two fold degenerate and
the explicit expressions are given in \cite{Hietanen:2009ex}.
The complete result from fermions and gauge fields is
\begin{equation}
S =  \sum_{i\ne j} \left\{
\ln \left[\sum_\mu \sin^2 \frac{1}{2}\left(\theta_\mu^i-\theta_\mu^j\right) \right]
- 4f
\ln \lambda_{n,o}( \theta^i-\theta^j+\phi)\right\} -4(N-1)f\ln\lambda_{n,o}(\phi).
\label{pertact}
\end{equation}

Independent of the actual values of $\theta_\mu^i$, the fermion
eigenvalues will have $(N-1)$ zero modes with periodic
boundary conditions when the mass is set to zero.
If all the $\theta_\mu^i$ are different for each $\mu$, then the
fermions should not have exact zero modes when we set
$p_\mu$ equal to $\left(\theta_\mu^i-\theta_\mu^j\right)$ with $i\ne j$. If the fermion spectrum
has more than $(N-1)$ zero modes, we will refer to the extra ones as doubler zero
modes.

In order to find the minimum of $S$, we consider the Hamiltonian
\begin{equation}
H = \frac{1}{2}\sum_{\mu,i} \left(\pi_\mu^i\right)^2 + \beta S.
\end{equation}
For large $\beta$, the Boltzmann measure $e^{-H}$ will be dominated
by the minimum of $S$. We can perform a HMC update of the $\pi,\theta$
system to find this minimum.

A choice for the order parameters associated with the $Z_N^4$ symmetries is~\cite{Bhanot:1982sh} 
\begin{equation}
P_\mu = \frac{1}{2} \left( 1 - \frac{1}{N^2}|\Tr U_\mu|^2\right)
=\frac{1}{N^2}\sum_{i,j}
\sin^2 \frac{1}{2}\left(\theta_\mu^i-
\theta_\mu^j\right) 
\end{equation}
The value $P_\mu=\frac{1}{2}$ corresponds to unbroken symmetry in $\mu$
direction and $P_\mu=0$ totally broken symmetry.

\subsection{Na\"ive fermions break the $Z_N^4$ symmetries}

\begin{figure}
  \centerline{\includegraphics[width=0.45\textwidth]{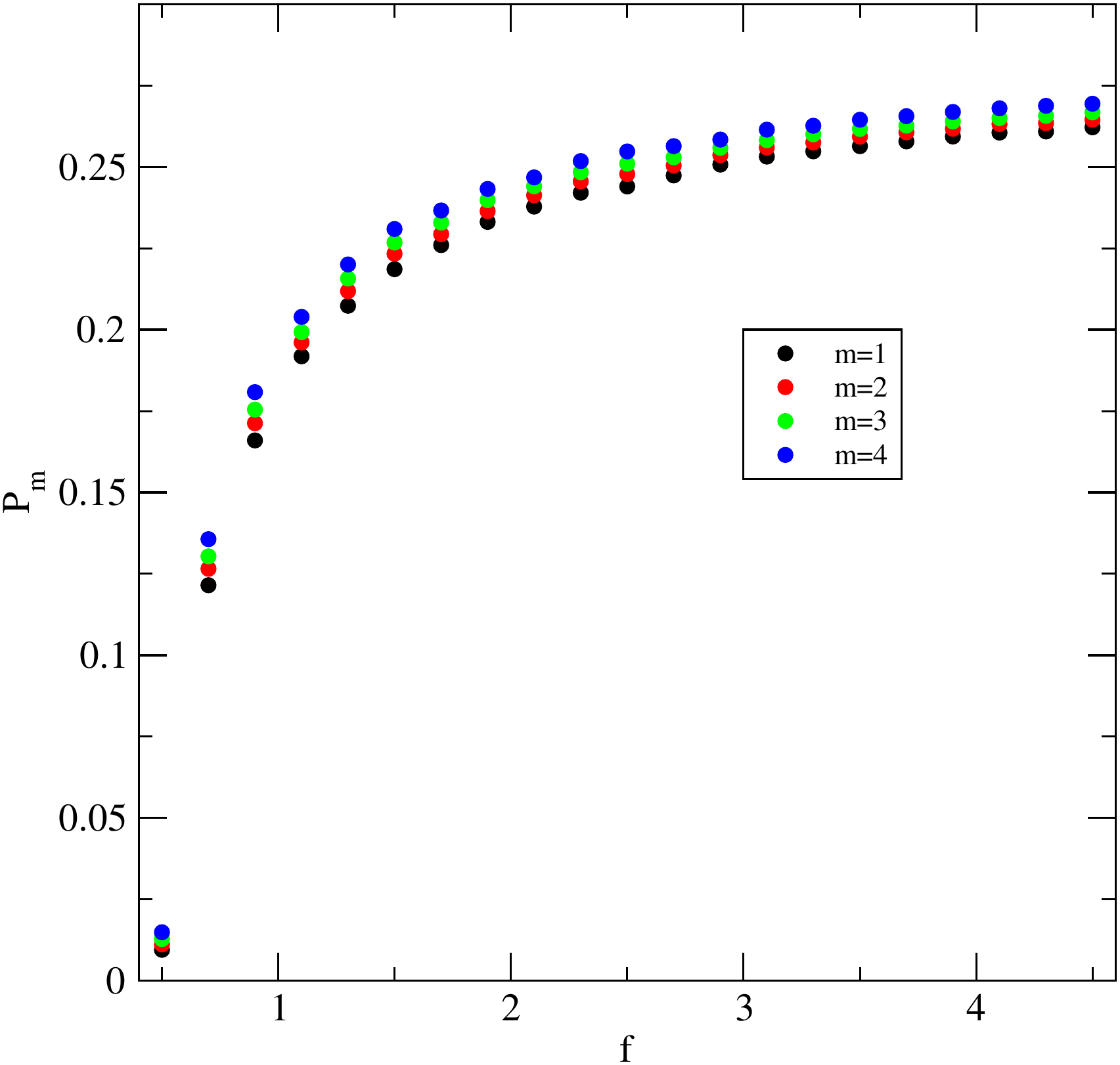}
    \includegraphics[width=0.45\textwidth]{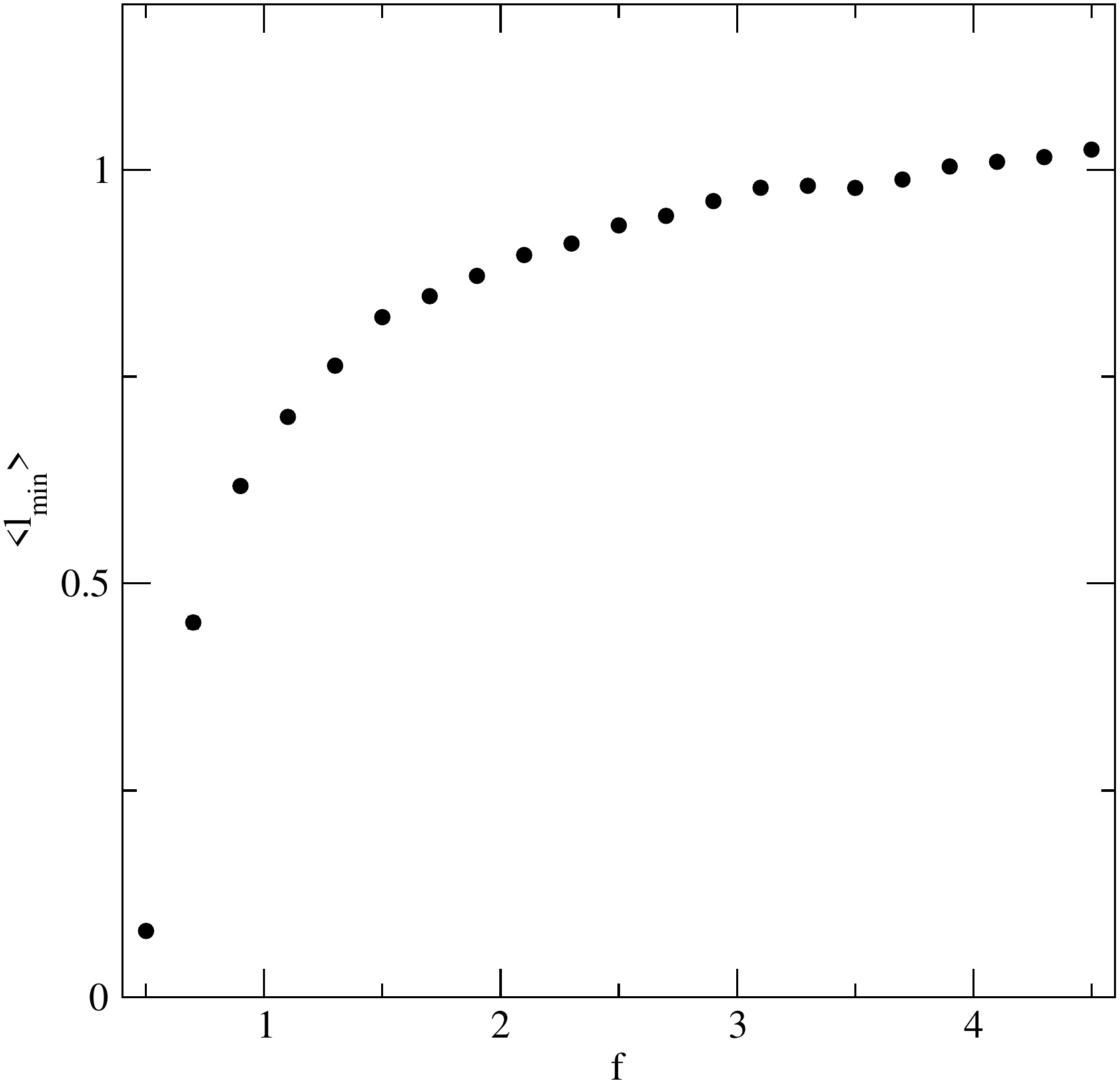}}
  \caption{Na\"ive fermions:
    Left: $P_\mu$ as a function of $f$ at $N=23$, $\beta=4$, and
    $\mu=0.01$. Right: Smallest eigenvalue, $\lambda(p)$ with
    $p_\mu=\left(\theta_\mu^i-\theta_\mu^j\right)$ and $i\ne j$, for
    $N=23$, $\beta=4$, and $\mu=0.01$.}
  \label{fig2}
\end{figure}

We assume periodic boundary conditions. We pick
one value of $N$ and $\beta$ and calculate $P_\mu$ 
as a function of $f$. In order to clearly see symmetry breaking we
use rotational symmetry on the lattice and
choose to label our directions such 
that $P_1< P_2 < P_3 < P_4$ for each configuration in our
thermalized ensemble. We set $\mu=0.01$ to avoid potential
singularities that could occur for the massless case. The plot for
$\beta=4$ confirms the breaking of the center symmetry left panel of 
Fig.~\ref{fig2}. The value is well below $\frac{1}{2}$.

With $N=23$, we expect $22$ exact zero modes 
for $\lambda(p)$ as explained in
section~\ref{wca}. We plot the average of the smallest 
eigenvalue, $\lambda(p)$ with $p_\mu=\left(\theta_\mu^i-\theta_\mu^j\right)$ 
and $i\ne j$, in right panel of Fig.~\ref{fig2}. 
It is clear that this eigenvalue is non-zero for
all values of $f$ indicating that there are no doubler zero modes.
The reason for the breaking of the $Z_N^4$ symmetries can be
understood by looking at the total action obtained from
(\ref{pertact}):
\begin{equation} S =  \sum_{i\ne j} 
\ln \left[\sum_\mu
\sin^2 \frac{1}{2}\left(\theta_\mu^i-
\theta_\mu^j\right) \right]
-2f \sum_{i\ne j} \ln \left[
\mu^2 + \sum_\mu \sin^2 (\theta^i_\mu-\theta^j_\mu)\right].
\end{equation}
The fermionic contribution cannot separate 
$\theta_\mu^i=\theta_\mu^j$ 
from $\theta_\mu^i=\theta_\mu^j+\pi$  
implying that the fermion
contribution alone will result in a distribution of eigenvalues
restricted to a width of $\pi$.

\subsection{Overlap fermions do not break the $Z_N^4$ symmetries}

\begin{figure}
\centerline{\includegraphics[width=0.45\textwidth]{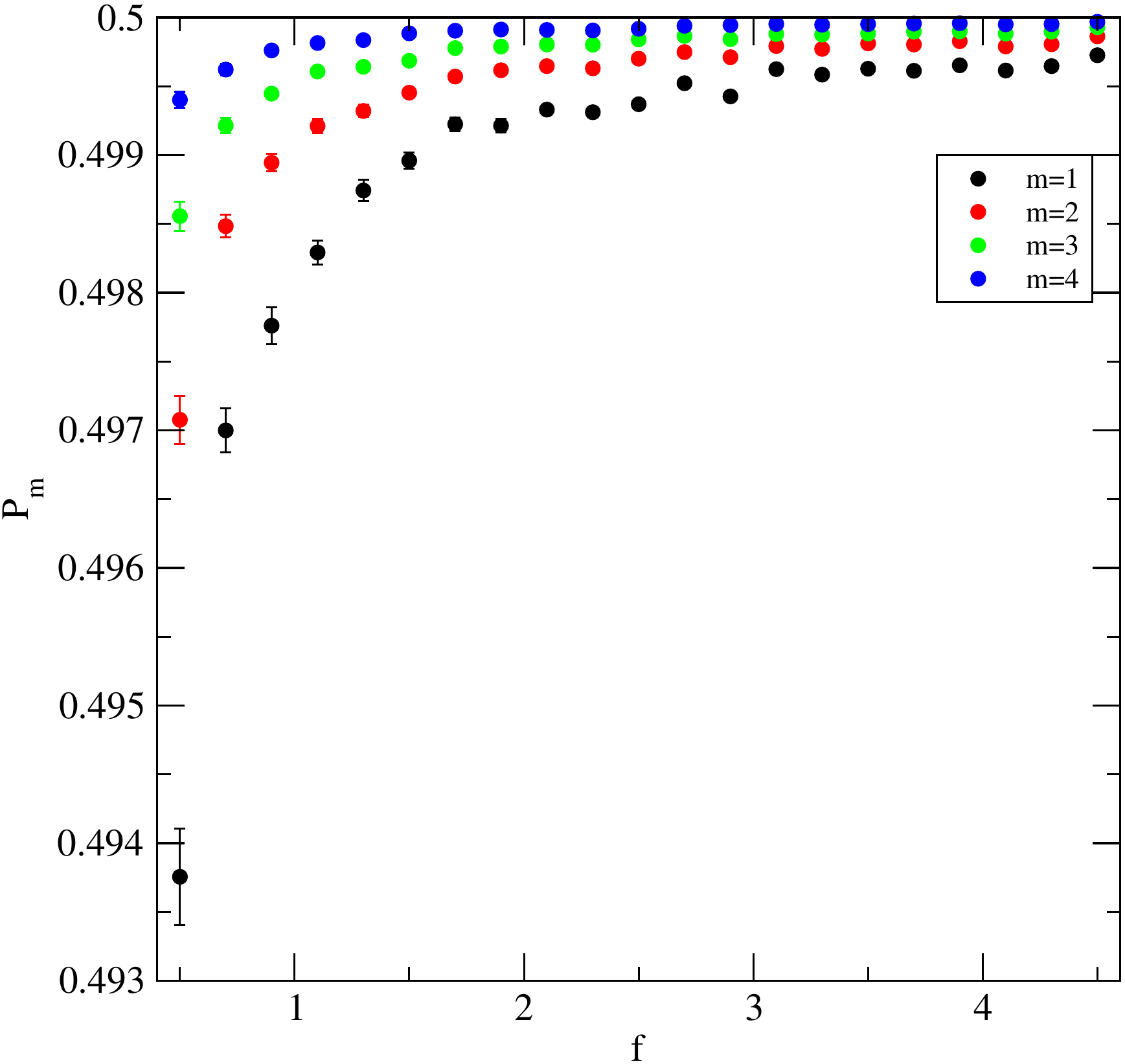}
\includegraphics[width=0.45\textwidth]{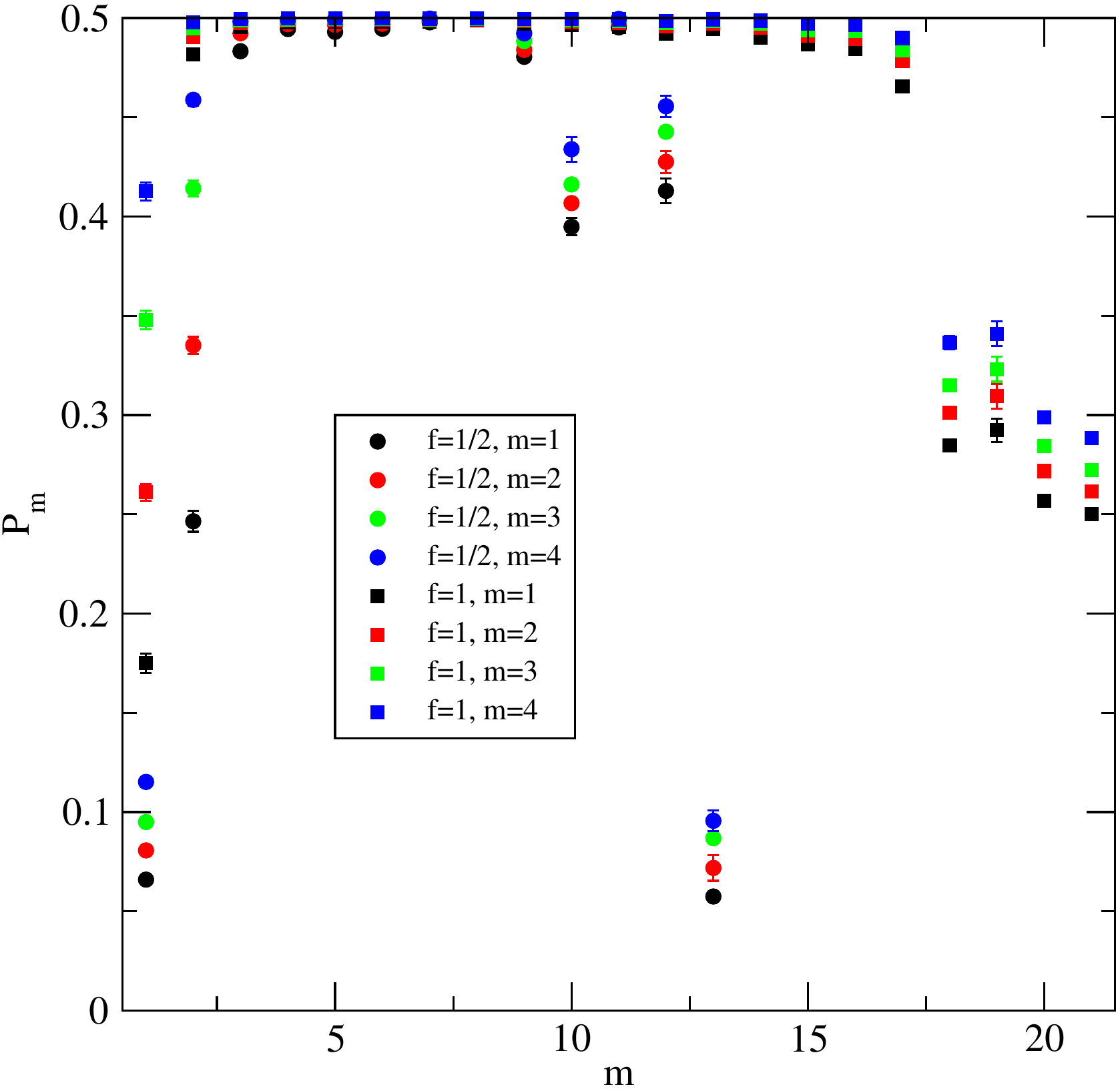}
}
\caption{Overlap fermions: Left:
Plot of $P_\mu$ as a function of $f$ at $N=23$, $\beta=1$, and
$\mu=0.01$. Right:Plot of $P_\mu$ as a function of $m$ at $N=23$, $\beta=1$, and
$\mu=0.01$ for two different values of $f$.
}
\label{fig4}
\end{figure}

In contrast to na\"ive fermions the overlap fermions do not break the
$Z_N^4$ symmetries. A plot of $P_\mu$ for several values of $f$
at $N=23$ and $\beta=1$ with $\mu=0.01$ in right panel of Fig.~\ref{fig4}
shows this to be the case. The very small deviation close to
$f=\frac{1}{2}$
is a consequence of finite $N$ effects. Since we probably are not able
to go beyond $N=23$ in the full simulation of the model, this
plot will serve as a guide to what one can expect in a full
simulation. 

\begin{figure}
\centerline{\includegraphics[width=0.45\textwidth]{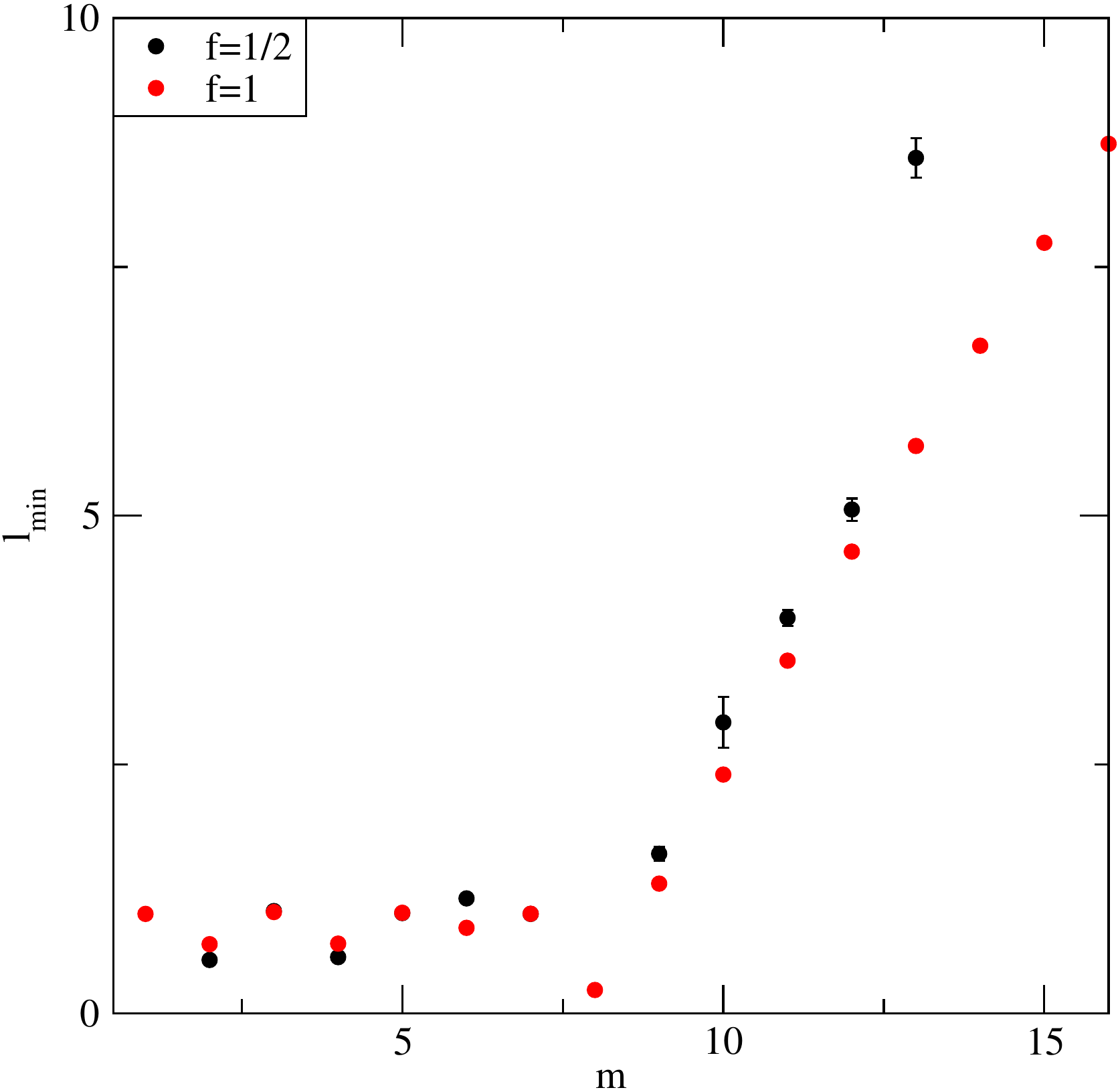}
\includegraphics[width=0.45\textwidth]{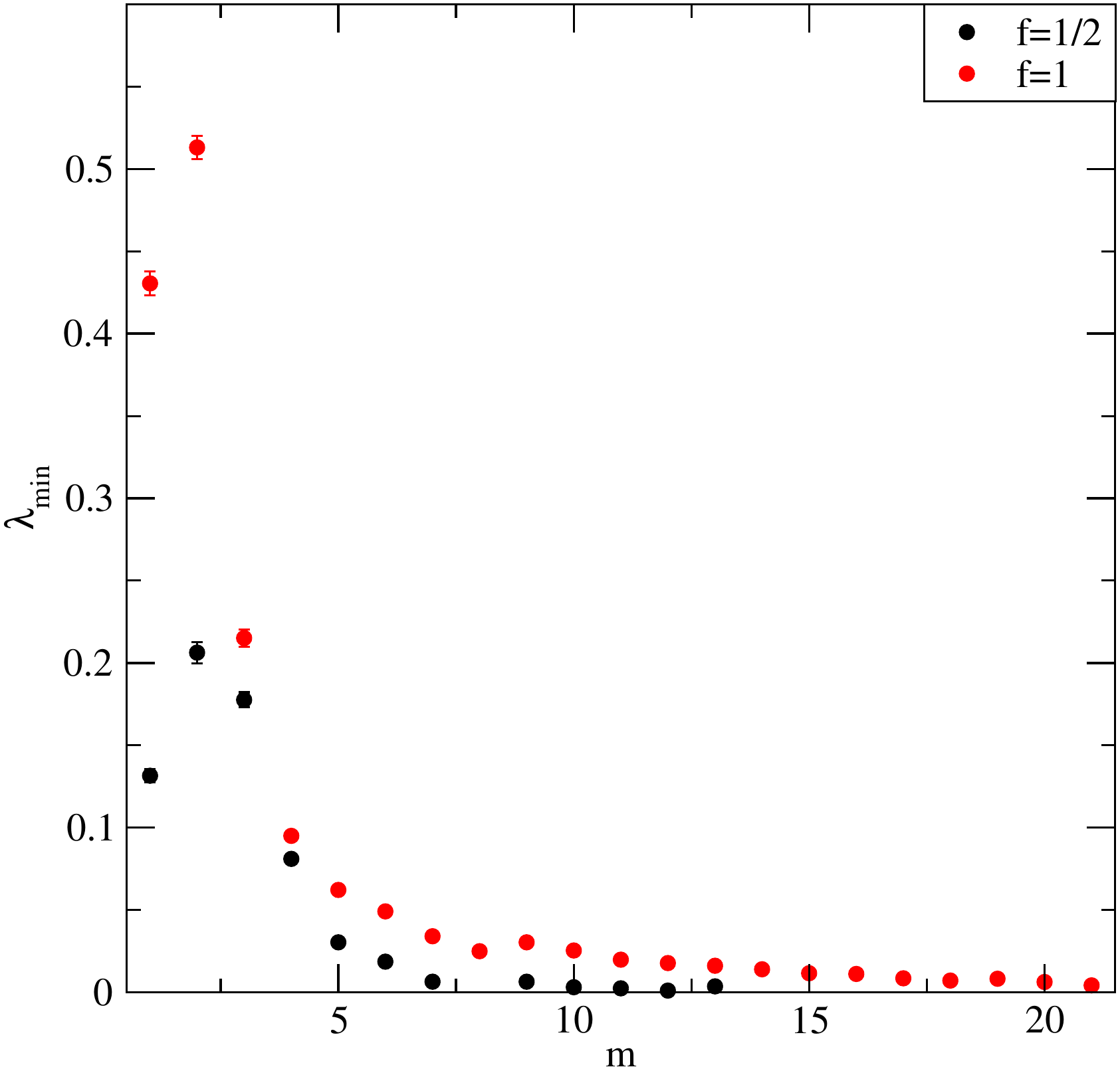}
}
\caption{Overlap fermions: Left
Plot of the lowest positive eigenvalue of $H$
as a function of $m$ at $N=23$, $\beta=1$, and
$\mu=0.01$ for two different values of $f$.
Right:Plot of  
smallest eigenvalue,
$\lambda(p)$ with $p_\mu=\left(\theta_\mu^i-\theta_\mu^j\right)$ 
and $i\ne j$, for
$N=23$, $\beta=1$, and
$\mu=0.01$.
}
\label{fig6}
\end{figure}

Because there are no doubler zero modes, we do not have any
restriction on the values for the Wilson mass, $m$, used in the
Wilson-Dirac kernel. However, we cannot make it arbitrarily large since
one can see by a direct computation that the large $m$ limit
of overlap fermions is na\"ive fermions~\cite{Narayanan:1994gw}.
A plot of $P_\mu$ as a function of $m$ is shown form $f=\frac{1}{2}$
and $f=1$ in right panel of Fig.~\ref{fig4}. It indicates that $3\le m
\le 8$ is an appropriate range of values of $m$ where the $Z_N^4$
symmetries are not broken for $f=\frac{1}{2}$ and that range only gets
bigger as $f$ increases. A plot of the lowest positive eigenvalue of
$H_w$ in left panel of Fig.~\ref{fig6} and the smallest eigenvalue,
$\lambda_o(p)$ with $p_\mu=\left(\theta_\mu^i-\theta_\mu^j\right)$,
$i\ne j$, in right panel of Fig.~\ref{fig6} shows that
there are no doubler zero modes in this range of $m$.
This range of $m$ can be used for full numerical simulation with
overlap fermions.

\subsection{Effect of fermion boundary conditions}
We have assumed periodic boundary conditions for fermions in the
previous two subsections. Other choices of boundary conditions 
that do not generate a U(1) anomaly amount
to replacing $V_\mu$ by $V_\mu e^{i\frac{2\pi k_\mu}{N}}$ with integer
valued $k_\mu$~\cite{Poppitz:2008hr}.
Physical results are expected to depend on the choice of boundary
conditions. This is in contrast to the case of large $N$ gauge
theories coupled to fundamental fermions. In that case, the
$e^{i\frac{2\pi k_\mu}{N}}$ factor can be absorbed by a change of gauge fields
that only changes the Polyakov loop and not the action. If the
$Z_N$ symmetries are not broken as is the case in the confined
phase, this change will not affect
physical results.

\begin{figure}
\centerline{\includegraphics[width=0.45\textwidth]{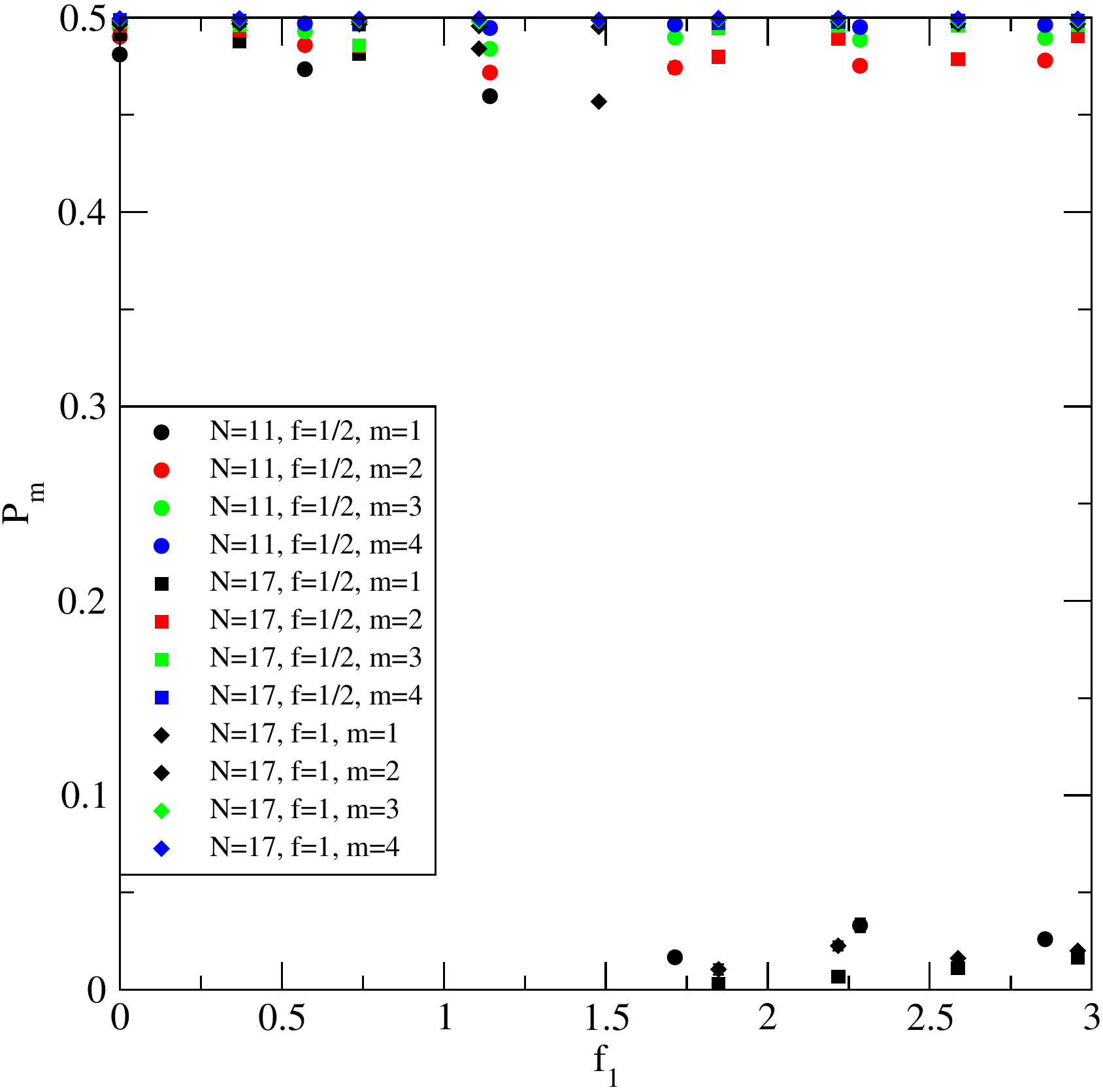}}
\caption{Overlap fermions:
Plot of $P_\mu$ as a function of $\phi_1$ for three different values
of $(f,N)$ with $\beta=1$ and
$\mu=0.01$.}
\label{fig8}
\end{figure}

In order to study the effect of boundary conditions on $Z_N^4$
symmetry breaking, we focus on $f=\frac{1}{2}$ and $f=1$.
We set $\phi_\mu=0$ for $\mu=2,3,4$ and varied $\phi_1$
by setting it to equal to $\frac{2\pi k}{N}$ with $k$ an
integer in the range $0\le k < N/2$. The plot of $P_\mu$
as a function of $\phi_1$ is shown in Fig.~\ref{fig8}. We see
that the $Z_N$ symmetry in the $\mu=1$ direction is broken
if $\phi_1 > \frac{\pi}{2}$. This seems to be the case in
the limit of large $N$ and seems to be roughly independent
of $f$. Furthermore, the $Z_N$ symmetries in the other three
directions with periodic boundary conditions are not broken.
This result could help us force feed momentum for quarks
in the adjoint representation. In order to pursue this, we
need to study the effect of $\phi_1$ on the chiral condensate
and see if chiral symmetry is restored when
the $Z_N$ symmetry is broken and if the chiral condensate
is independent of the value of $\phi_1$ when the $Z_N$ symmetry
is not broken.

\section{Discussion}
We have provided perturbation theory arguments that the $Z_N^4$
symmetry is restored in the large $N$ limit for overlap fermions but
not for na\"ive fermions. The details of perturbation theory
calculations with framework to HMC-lattice simulations are given in
\cite{Hietanen:2009ex}.  The natural continuation is to perform
lattice calculations on several $N$ and calculate physical
observables, e.g., string tension and chiral condensate. 

\acknowledgments

The authors acknowledge partial support by the NSF under grant number
PHY-0854744. AH also acknowledges partial support by the U.S. DOE
grant under Contract DE-FG02-01ER41172.

\end{document}